\documentclass[fleqn,usenatbib,twocolumn]{mnras}

\usepackage{color}

\def\comment#1{}

\usepackage{newtxtext,newtxmath}

\usepackage[T1]{fontenc}

\DeclareRobustCommand{\VAN}[3]{#2}
\let\VANthebibliography\thebibliography
\def\thebibliography{\DeclareRobustCommand{\VAN}[3]{##3}\VANthebibliography}

\usepackage{graphicx}	
\usepackage{amsmath}	

\usepackage{amssymb}	

\title[Tests of weak equivalence principle]{Tests of weak equivalence principle with the gravitational wave signals in the LIGO-Virgo catalogue GWTC-1}

\author[S.-C. Yang et al.]{Shu-Cheng Yang$^{1,2}$, Wen-Biao Han,$^{1,2,3,4,5}$ 
\thanks{E-mail: wbhan@shao.ac.cn} and Gang Wang$^{1}$
\\
$^{1}$Shanghai Astronomical Observatory, Chinese Academy of Sciences, Shanghai 200030, China\\
$^{2}$School of Astronomy and Space Science, University of Chinese Academy of Sciences,  Beijing 100049, China\\
$^{3}$School of Fundamental Physics and Mathematical Sciences, Hangzhou Institute for Advanced Study, UCAS, Hangzhou 310024, China\\
$^{4}$International Centre for Theoretical Physics Asia-Pacific, Beijing/Hangzhou, China\\
$^{5}$Key Laboratory for Research in Galaxies and Cosmology, Shanghai Astronomical Observatory, Shanghai 200030, China\\
}

\date{Accepted XXX. Received YYY; in original form ZZZ}

\pubyear{2015}

\begin{document}
\label{firstpage}
\pagerange{\pageref{firstpage}--\pageref{lastpage}}
\maketitle

\begin{abstract}
The weak equivalence principle (WEP) is the cornerstone of gravitational theories. At the local scale, WEP has been tested to high accuracy by various experiments. On the intergalactic distance scale, WEP could be tested by comparing the arrival time of different messengers emitted from the same source. The gravitational time delay caused by massive galaxies is proportional to $\gamma+1$, where the parameter $\gamma$ is unity in general relativity. The values of $\gamma$ for different massless particles should be different if WEP is violated, i.e., $\Delta \gamma$ is used to indicate the deviation from WEP. So far, $|\Delta \gamma|$ has been constrained with gamma-ray bursts, fast radio bursts, etc. Here we report a new constraint of $|\Delta \gamma|$ by using the gravitational wave data of binary black hole coalescences in the LIGO-Virgo catalogue GWTC-1. The best constraints imply that $|\Delta \gamma| \lesssim  10^{-15}$ at 90 per cent confidence level. 

\end{abstract}

\begin{keywords}
gravitation - gravitational waves - methods: data analysis
\end{keywords}



\section{Introduction}

%

Weak equivalence principle (WEP) assumes that all freely falling bodies at the same space-time points will undergo the same acceleration in a given gravitational field, independent of their properties and rest mass. WEP, which is rather fundamental in physics, has withstood numerous experimental tests at the local scale\citep{will2014confrontation}. A measurement on the fractional difference in acceleration between two bodies is $\mathrm{E\ddot{o}tv\ddot{o}s}$ ratio $\eta = 2|a_{1} - a_{2}|~/~|a_{1} + a_{2}|$, where $a_1$ and $a_2$ are the free-fall accelerations of the two bodies \citep{will2014confrontation}. To date, the bounds on $\eta$ have reached levels of $10^{-15}$ by MICROSCOPE's satellite experiments \citep{touboul2017microscope}.

Gravitational field can lead to an extra time delay on the propagation of photons and gravitational waves(GWs) which is called as Sharpio time delay \citep{shapiro1964fourth}. This delay is proportional to $\gamma+1$, where $\gamma$ is the parameterized post-Newtonian parameter ($\gamma  = 1$ in GR(general relativity)) \citep{will2014confrontation}. If WEP is violated, different massless particles will have different values of $\gamma$ when they freely fall in the gravitational field. For instance, the values of $\gamma$ for signal 1 and signal 2 are $\gamma_1$ and $\gamma_2$, respectively, then $|\Delta \gamma|$ ($|\gamma_{1} - \gamma_{2}|$) can be used to quantitatively represent the derivation of WEP. There shall be $|\Delta \gamma| = 0$ in the case of GR.   

Usually, this gravitational time delay is too small to be used to test WEP. However, in astrophysics, due to the huge mass of galaxies and super long distance of sources, it offers a unique opportunity to test WEP on the intergalactic scale. According to WEP, freely falling massless particles emitted from the same astrophysical source shall follow an identical geodesic and experience the same Shapiro time delay that caused by the presence of a gravitational potential. Considering two different signals emitted from the same source, the observed time-delays  ($\Delta t_{obs}$) could be greater than the difference of the Shapiro time delay of the two particles $\Delta t_{\rm gra}$, which is proportional to $|\Delta \gamma|$. Therefore, people could give an upper limit on $|\Delta \gamma|$  by observing $\Delta t_{obs}$ of different signals emitted from the same source \citep{krauss1988test}. To date, several constraints on $|\Delta \gamma|$ have obtained using different astrophysical events, including emissions from supernova event SN1987A\citep{krauss1988test, longo1988new}, gammy-ray bursts\citep{gao2015cosmic} and fast radio bursts (FRBs) \citep{wei2015testing, tingay2016limits, xing2019limits}.  Recently, GWs have also been used to test WEP\citep{kahya2016djconstraints, wu2016testing,wei2017multimessenger},  the constraints on $|\Delta \gamma|$ obtained from GWs is $10^{-9}$. By now, Wei et al. gave a constraint of $10^{-13}$ for different particles (photons and neutrinos)\citep{wei2016limits}. Xing et al. gave a constraint of $10^{-16}$ for the same particles (photons) with different energies\citep{xing2019limits} 

The aforementioned methods took $\Delta t_{\mathrm{obs}}$ between low-frequency and high-frequency GWs or EM waves (or time delay between different messengers) as the maximum possible value of time delay due to the violation of WEP. Because they did not remove the intrinsic time delay when these signals emitted (i.e. $\Delta t_{\mathrm{e}}$), they obtained rough upper limits on $|\Delta \gamma|$ \citep{wu2016testing,wei2017multimessenger,yao2019new}. 

Fortunately, we can theoretically calculate the inspiral and merger of binary black holes(BBHs), then the intrinsic time delay between the low and high-frequency GWs in one event could be modelled accurately. This supplies us an opportunity to remove the intrinsic time delay and estimate the up-limit of $|\Delta \gamma$| better from the GW data, and what we need is the waveform templates of GW that including $\Delta \gamma$. The violation of WEP will contribute $\Delta t_{\rm gra}$ for GWs with different frequencies during the signals propagating through the galaxies, which will cause the dephasing from the waveforms predicted by GR. 

In this work, due to the details of the host galaxies of these events are unclear, we ignore the time delay due to these host galaxies. In addition, the violation of WEP will induce the modification of gravitational theory, then the generated waveforms at the merger should be different from GR. However, this modification should be very small considering the current constraints on the WEP in the literature (especially compared to the propagation effect of GWs). For simplicity, we ignored the influence of WEP violation during the generation of GWs. The dephasing of waveforms is introduced when gravitational waves passing through the Galaxy. We construct a new gravitational waveform template by adding a modification term (see Section~\ref{sec:method}) on the IMRPhenomPv2 waveforms\citep{hannam2014simple, husa2016frequency, khan2016frequency}, which is also used in LIGO-Virgo's parameter estimation\citep{abbott2019gwtc}.Then we employ this new template to estimate the parameter $\Delta \gamma$ by using GW events in GWTC-1 by a Bayesian inference software named BILBY\citep{ashton2019bilby}. 
This Letter is organized as follows. Section~\ref{sec:method} introduces our methodology to test the WEP. In Section~\ref{sec:results} we demonstrate our results with 10 GW events. The conclusion and discussion are given in Section~\ref{sec:discussion}.

\section{Method}
\label{sec:method}
If WEP is violated, GWs with different frequencies may experience different Shapiro time delays \citep{shapiro1964fourth}. To give a constraint on the violation of WEP, we assume all the uncertainty of arrival time of GW is caused by the violation of WEP, which could be described by the difference of the Shapiro time delay of the two particles $\Delta t_{\rm gra}$. For the same source, considering GWs emitted at $t_{\rm{e}}$ and $t'_{\rm{e}}$ with different frequency, which will be received at corresponding arrival times $t_{\rm{a}}$ and $t'_{\rm{a}}$. If the difference of emitting time ($\Delta t_{\rm{e}} = t_{\rm{e}} - t'_{\rm{e}}$) is so little that the cosmological inflation effect could be ignored, then the delay of arrival times of the two GWs  ($\Delta t_{\rm{a}} = t_{\rm{a}} - t'_{\rm{a}}$) is 

\begin{align}
\Delta t_{\mathrm{a}} =  \left(1 + z\right) \Delta t_{\rm e} + \Delta t_{\rm gra},
\label{eqn:timedelay} 
\end{align}
where $z$ is the cosmological redshift, and $\Delta t_{\rm gra}$  would be\citep{wei2015testing}
\begin{align}
\Delta t_{\rm gra} = \frac{\Delta \gamma}{c^{3}}\int_{{r_{\rm o}}}^{{r_{\rm e}}} U\left({r}\right)d{r},
\label{eqn:t_gra} 
\end{align}
where $\Delta \gamma$ ($= \gamma - \gamma'$) may be negative or non-negative, $r_{\rm o}$ and $r_{\rm e}$ are the locations of observation and the source of GWs, $U\left(r\right)$ denotes the gravitational potential. For simplicity, in this work, we only consider the gravitational potential caused by Milky Way, and $\Delta t_{\rm gra}$ would be \citep{wu2016testing}
\begin{align}
\Delta t_{\rm gra} = \Delta \gamma \left[\frac{GM_{\mathrm{MW}}}{c^{3}} \mathrm{ln}\left(\frac{\left[d + (d^{2} - b^{2})^{1/2}\right] \left[r_{G} + s_{n}\left(r_{G}^{2}-b^{2} \right)^{1/2} \right]}{b^2}\right) \right],
\label{eqn:t_gra2} 
\end{align}
where the Milky Way mass $M_{\mathrm{MW}} \approx 6 \times 10^{11}~M_{\odot}$, $d$ denotes the distance from the source to the Milky Way centre, $b$ represents the impact parameter of the GW paths relative to the centre of Milky Way, and the distance from the Sun to the centre of Milky Way $r_{G} \approx 8 ~ \mathrm{kpc}$, $s_n = + 1$ denotes the source located along the direction of the Milky Way centre, and  $s_n = -1$ denotes the source located along the direction that is pointing away from the Milky Way centre. The positions of sources usually use celestial coordinates(right ascension $\beta$ and declination $\delta$), therefore we use a transform formula\citep{yao2019new} to convert celestial coordinate to $b$. 

The Shapiro time delay difference $\Delta t_{\rm gra}$ for one GW event in different frequencies could cause the dephasing of the waveforms comparing with GR's templates. Therefore, we need a modified waveform template to include this effect. Our modified waveform in the frequency domain is
\begin{align}
\tilde{h}(f) = \tilde{A}(f) e^{i \left[\Psi_{\rm GR}(f) + \delta\Psi(f) \right]},
\label{eqn:waveform} 
\end{align}
where $\tilde{A}(f)$ denotes the complex amplitude, $\Psi_{\rm GR}(f)$ denotes the complex phase that predicted by GR, and $\delta\Psi(f)$ is the modification term produced by the deviation of WEP. In this work, inspired by the gravitational waveforms with time-delay-based dephasing\citep{mirshekari2012constraining, will1998bounding} and Eqn.~\ref{eqn:t_gra2},  we finally work out the modification term
\begin{align}
\begin{split}
\delta \Psi(f) &= \frac{\pi\Delta \gamma}{\Delta f} \left[\frac{GM_{\mathrm{MW}}}{c^{3}} \mathrm{ln}\left(\frac{\left[d + (d^{2} - b^{2})^{1/2}\right] \left[r_{G} + s_{n}\left(r_{G}^{2}-b^{2} \right)^{1/2} \right]}{b^2}\right) \right] \\
&\cdotp \left(1 + z\right )^{2} {f}^{2},
\label{eqn:delta_phi} 	
\end{split}
\end{align}
where $\Delta f = f - f'$, and $f, ~f'$ is two different frequencies of GWs in one event. The violation of WEP changes the free-fall of particles in an external gravitational field. Intuitively, the modification of free-fall should be proportional to the energy (mass) of the particle. Considering $E = h f$ (where $h$ is the Plank constant), we have the assumption that $\Delta \gamma \propto \Delta f$, then we use the waveform model equation~(\ref{eqn:waveform}) to estimate parameters of GW events in GWTC-1.

In this work, we use a Bayesian parameter estimation software named BILBY\citep{ashton2019bilby} to estimate the parameters with the above gravitational waveform templates. In Bayesian parameter estimation, the prior distributions of different parameters of GW source models are needed first. For parameters except $\Delta \gamma$, we use BILBY's default parameter priors for BBHs\citep{ashton2019bilby}. For $\Delta \gamma$, we introduce a modified logarithmic prior that could cover both the negative and non-negative value. The modified logarithmic prior is described by
\begin{align}
\Delta \gamma(\alpha) = \left\{
\begin{aligned}
10^{-\frac{1}{\alpha}} ~ &(\alpha > 0) \\
0 ~ &(\alpha = 0)\\
-10^{\frac{1}{\alpha}} ~ &(\alpha < 0)
\end{aligned}
\right.
\label{eqn:inverse_log} 
\end{align}
where $\alpha$ is an uniform distribution parameter, then the prior of $\Delta \gamma$  covers from the negative to positive value continuously.

\section{Results}
\label{sec:results}

\begin{figure}
\begin{center}
\includegraphics[height=2.2in]{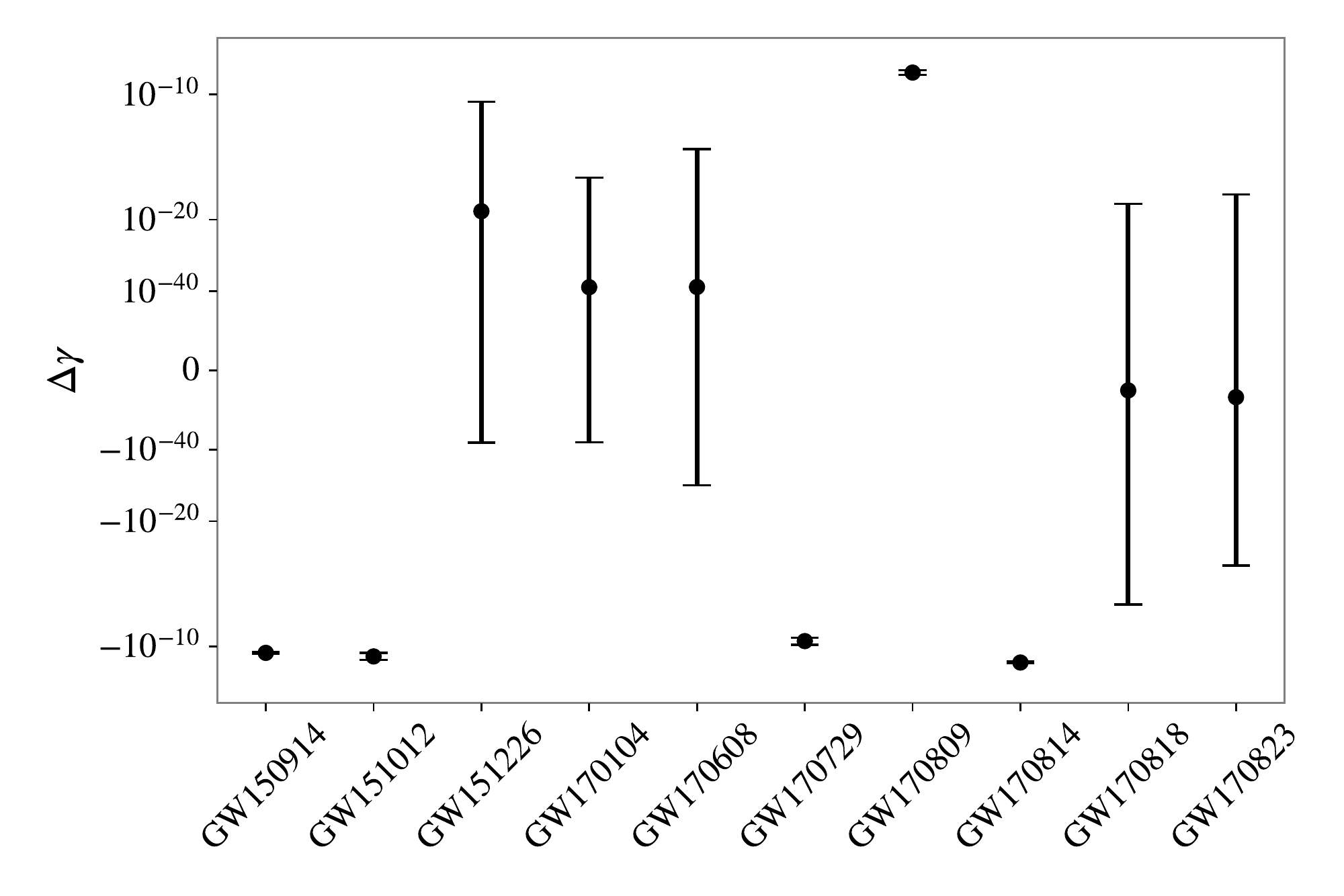}
\caption{The posterior distribution of $\Delta \gamma$ for GW events of BBH in GWTC-1 (90 per cent confirmation level). A modified logarithmic prior (See Section~\ref{sec:method}), which could cover both the negative and non-negative value, is assumed on the prior distribution of $\Delta \gamma$. $\Delta \gamma$ belongs to $[-1.15\times 10^{-5}, 1.15\times 10^{-5}]$.  }  
\label{fig:r1}
\end{center}
\end{figure}

\begin{figure}
\begin{center}
\includegraphics[height=1.9in]{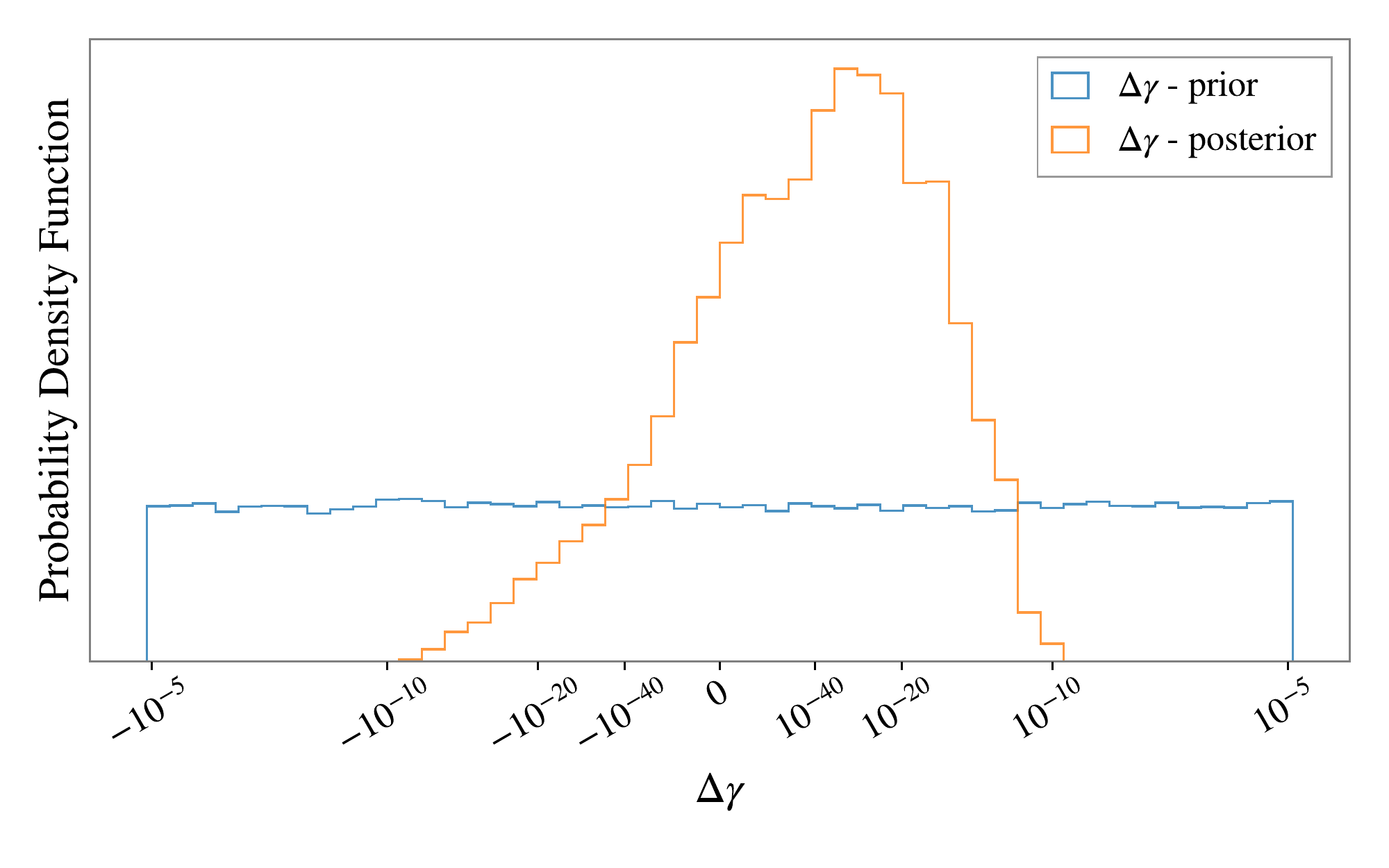}
\includegraphics[height=1.9in]{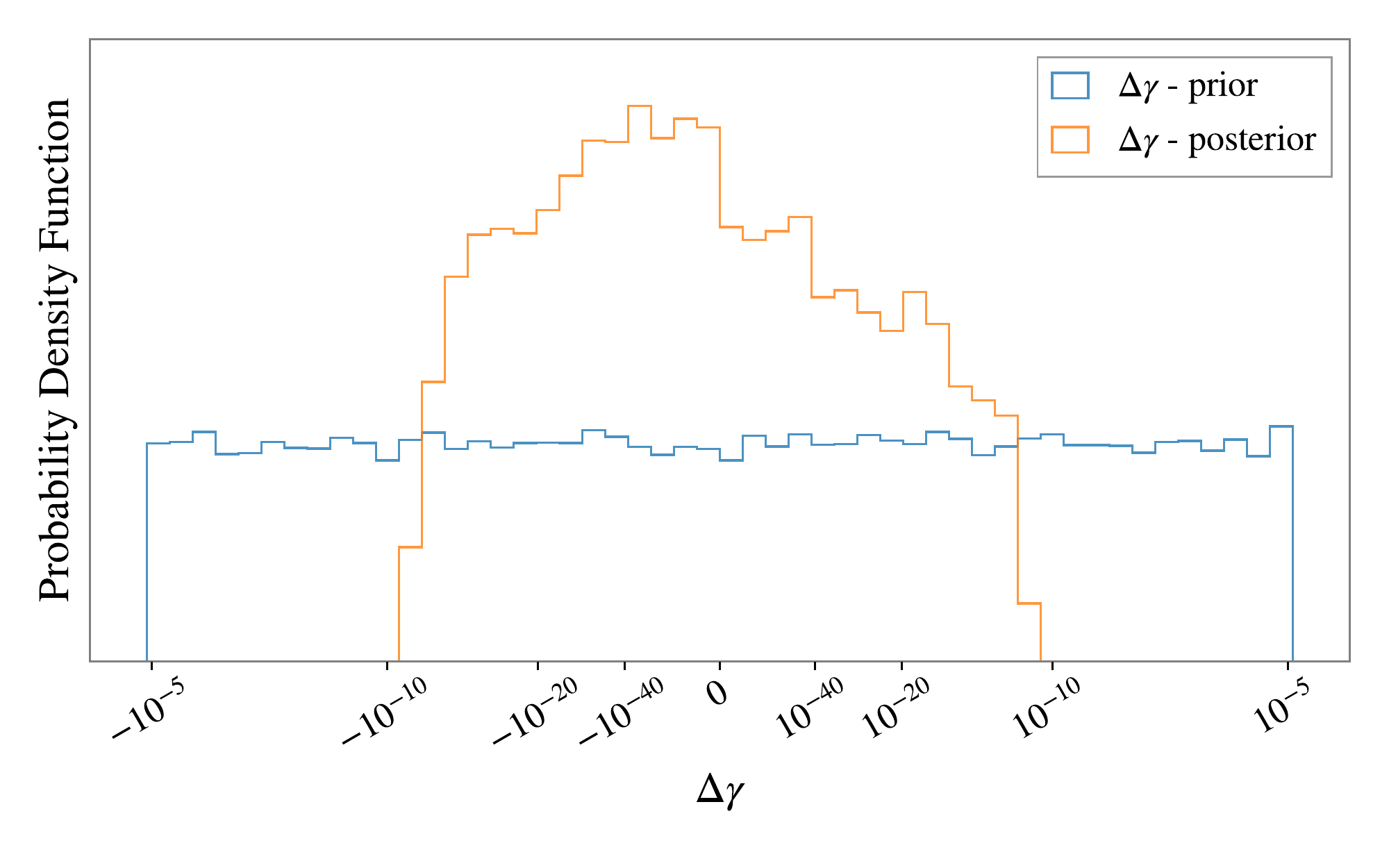}
\caption{The prior and posterior distribution of $\Delta \gamma$ for GW170104 (top) and GW170823 (bottom). A modified logarithmic prior (See Section~\ref{sec:method}), which could cover both the negative and non-negative value, is assumed on the prior distribution of $\Delta \gamma$, which belongs to $[-1.15\times 10^{-5}, 1.15\times 10^{-5}]$.}
\label{fig:r3}
\end{center}
\end{figure}

\begin{figure}
\begin{center}
\includegraphics[height=2.2in]{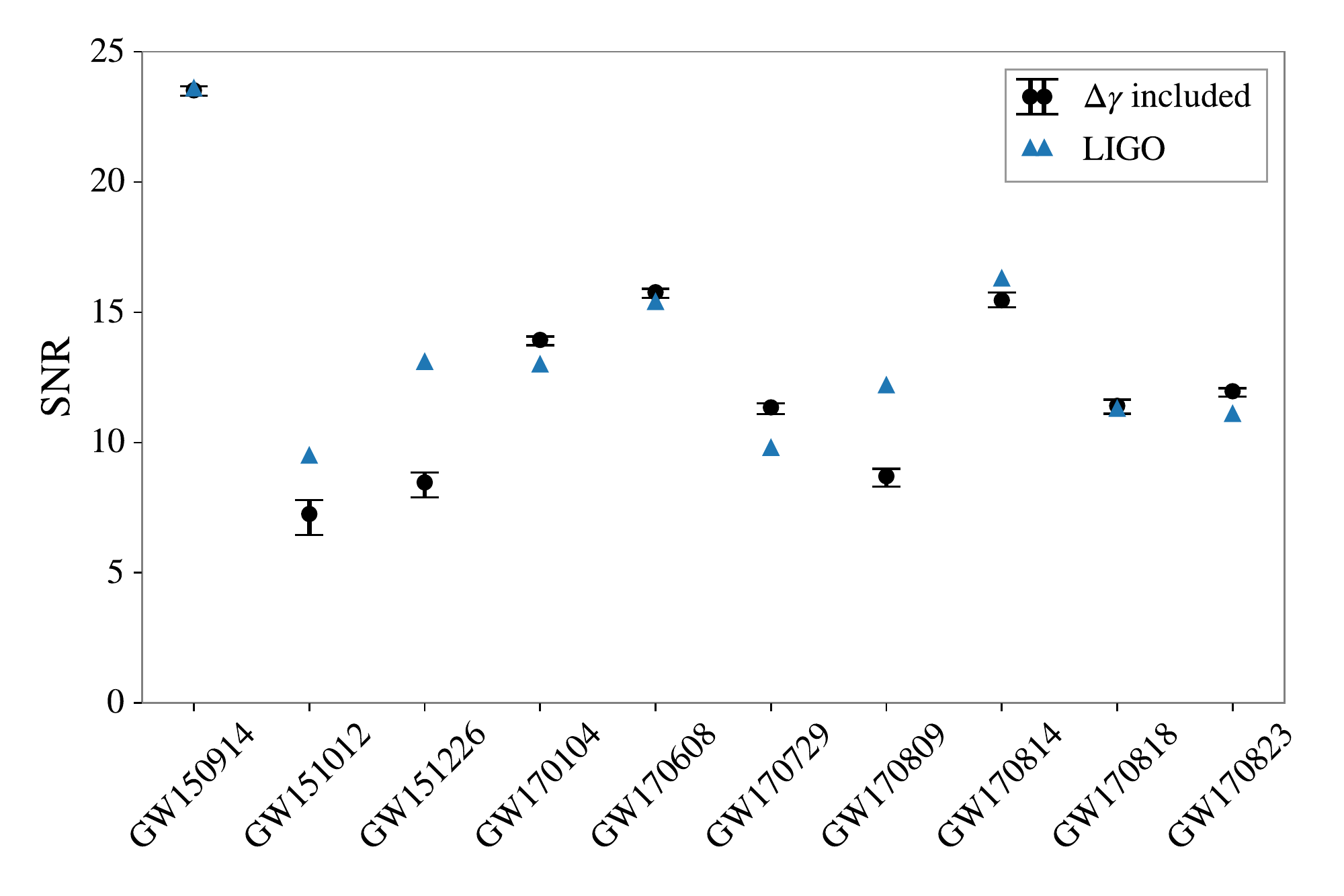}
\caption{The distribution of network SNR for GW events of BBH in GWTC-1 (90 per cent confirmation level). For points in black, a modified logarithmic prior (See Section~\ref{sec:method}), which could cover both the negative and non-negative value, is assumed on the prior distribution of $\Delta \gamma$. Blue points show the corresponding network SNR of the LIGO-Virgo Collaboration (See table~1. in \citet{abbott2019gwtc}). }  
\label{fig:r2}
\end{center}
\end{figure}

\begin{table}\scriptsize
	\centering
	\caption{The posterior distribution of $\Delta \gamma$ (modified logarithmic prior, see Section~\ref{sec:method}), correspondent SNR and up-limit of $|\Delta \gamma|$ for BBH GW events in GWTC-1.}
	\label{tab:table_1}
	\begin{tabular}{lccccccr} 
		\hline
		Events &  Posterior Distribution of $\Delta \gamma$  &  SNR  & Up-limit of $|\Delta \gamma|$\\
		\hline

		GW150914 & $-1.9\times 10^{{-10}_{-0.0}^{+0.0}}$  & $23.52_{-0.21}^{+0.16}$ & $|\Delta \gamma| \leq 1.9\times 10^{{-10}_{-0.0}^{+0.0}}$ \\	

		GW151012 & $-2.7\times 10^{{-10}_{-0.2}^{+0.1}}$ & $7.26_{-0.81}^{+0.53}$ & $|\Delta \gamma| \leq 2.7\times 10^{{-10}_{-0.2}^{+0.1}}$ \\

		GW151226 & $-8.3\times 10 ^{-47} \leq \Delta \gamma \leq 4.7\times 10 ^{-11}$  & $8.47_{-0.58}^{+0.39}$ & $|\Delta \gamma| \leq 4.7\times 10^{-11}$\\

		GW170104 & $-7.1 \times 10^{-45} \leq \Delta \gamma \leq 6.2 \times 10^{-16}$  & $13.94_{-0.20}^{+0.13}$ & $|\Delta \gamma| \leq 6.2\times 10^{-16}$\\

		GW170608 & $-1.2\times 10^{-27} \leq \Delta \gamma \leq  1.1\times 10^{-13}$  & $15.77_{-0.21}^{+0.13}$ & $|\Delta \gamma| \leq 1.1\times 10^{-13}$\\

		GW170729 & $-5.9\times 10^{{-11}_{-0.3}^{+0.1}}$  & $11.35_{-0.25}^{+0.15}$  & $|\Delta \gamma| \leq 5.9\times 10^{{-11}_{-0.3}^{+0.1}}$\\

		GW170809 & $7.8\times 10^{{-10}_{-0.1}^{+0.1}}$  & $8.71_{-0.40}^{+0.28}$ & $|\Delta \gamma| \leq 7.8\times 10^{{-10}_{-0.1}^{+0.1}}$\\

		GW170814 & $-4.7\times 10^{{-10}_{-0.0}^{+0.0}}$  &  $15.46_{-0.27}^{+0.30}$ & $|\Delta \gamma| \leq 4.7\times 10^{{-10}_{-0.0}^{+0.0}}$\\

		GW170818 & $-7.1\times 10^{-13} \leq \Delta \gamma \leq 1.3 \times 10^{-18}$  &  $11.41_{-0.30}^{+0.24}$ & $|\Delta \gamma| \leq 7.1\times 10^{-13}$\\

		GW170823 & $-1.0 \times 10^{-15} \leq \Delta \gamma \leq 1.4 \times 10^{-17}$  & $11.97_{-0.20}^{+0.12}$ & $|\Delta \gamma| \leq 1.0\times 10^{-15}$\\
		\hline
	\end{tabular}
\end{table}

Now we demonstrate our results using the above methods, and our data is based on the 10 BBH GW events in GWTC-1. We must emphasize that our method in principle can not measure the value of $|\Delta \gamma|$. This is because we assume all the errors of observation are due to the violation of WEP, then what we get here is the up-limit of $|\Delta \gamma|$, i.e., $|\Delta \gamma|$ should be less than the estimations we obtained in this letter. 

Fig.~\ref{fig:r1} shows the posterior distribution  of $\Delta \gamma$  (90 per cent confirmation level). Here, the prior of $\Delta \gamma$ obeys the prior distribution given in Section~\ref{sec:method}, which could cover both the negative and non-negative $\Delta \gamma$ values continuously, i.e., $\Delta \gamma \in [-1.15\times 10^{-5}, 1.15\times 10^{-5}]$.  Fig.~\ref{fig:r3} shows the prior and posterior distributions of $\Delta \gamma$ for GW170104 and GW170823, which have the most stringent constraints in GWTC-1. Figure.~\ref{fig:r2} displays the distribution of network SNR (matched-filter SNR) of our results, and the SNR(Signal-to-Noise Ratio) results\citep{abbott2019tests} of LIGO-Virgo are also plotted. The detailed information of  Figs ~\ref{fig:r1} and \ref{fig:r2} and the final up-limits of $|\Delta \gamma|$ are shown in Table.~\ref{tab:table_1}.

In Fig.~\ref{fig:r1}, the results can be divided into three different groups about $\Delta \gamma$'s posterior distribution. For group~1(GW150914, GW151012, GW170729, GW170814), the constraint is $\Delta \gamma \gtrsim -10^{-10}$. For group~2 (GW170809), the result shows $\Delta \gamma \lesssim 10^{-10}$. For group~3 (GW151226, GW170104, GW170608, GW170818, GW170823), $\Delta \gamma$ is constrained more stringently. Attending that the error bar of the group~3 seems to be wider than the group~1 and the group~2. This is because the coordinates of $\Delta \gamma$ is scaled in Fig.~\ref{fig:r1}. Actually, group~3 has even smaller error-bar than the group~1 and group~2. From these results, we get the up-limits of $\Delta \gamma$. As shown in Table~\ref{tab:table_1}, the most stringent constraints come from GW170104 and GW170823, i.e. $|\Delta \gamma| \lesssim 10^{-15}$. Because the estimate for each event 
just requires $\Delta \gamma$ can not be larger than a certain value, we then choose the constraints of GW170104 and GW170823 as the final up-limits we estimated for $\Delta \gamma$. 

\begin{figure}
\begin{center}
\includegraphics[height=3.6in]{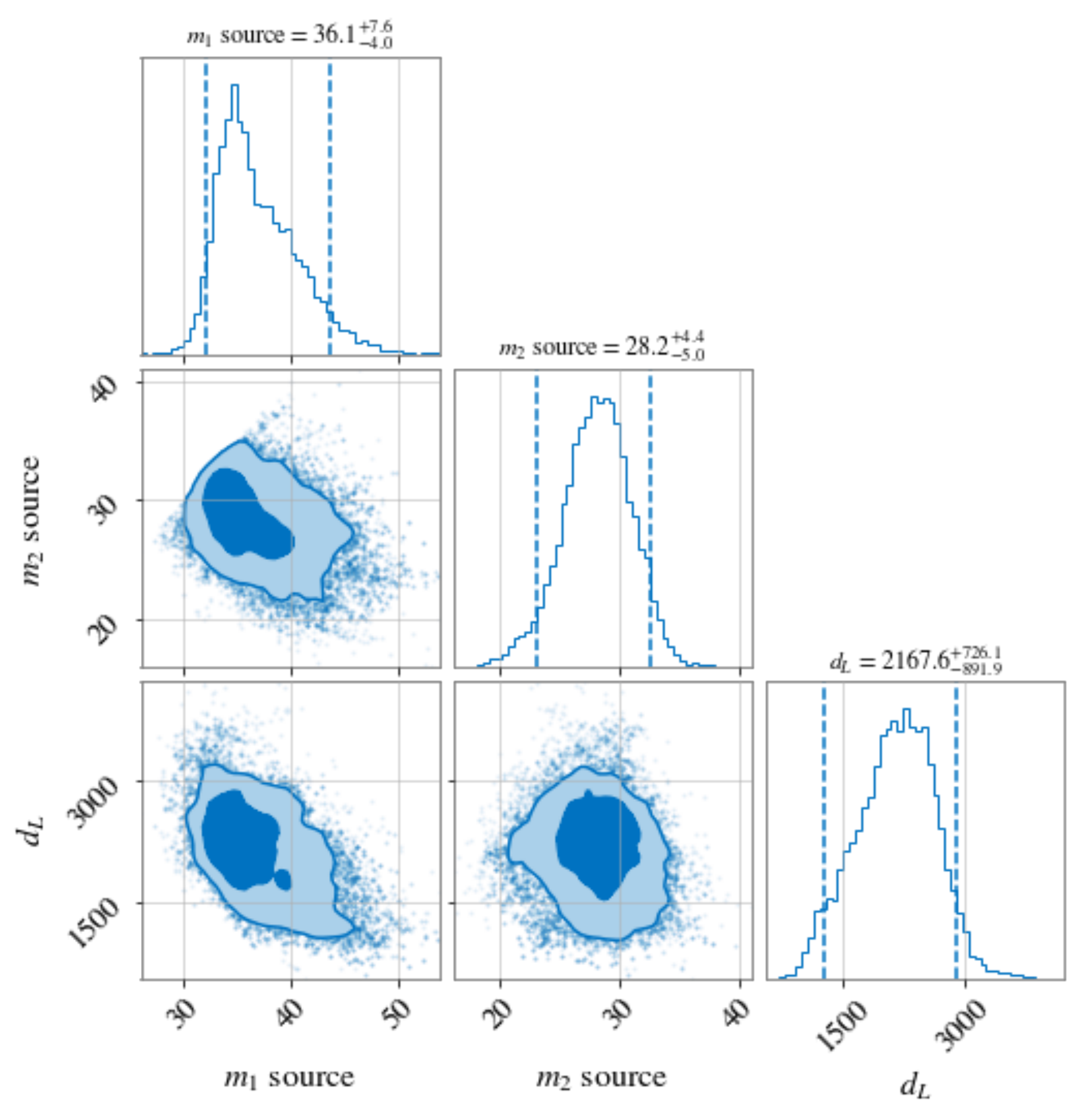}
\caption{The posterior probability density for binary component masses (in solar mass) in the source frame, luminosity distance (in Mpc) of GW170823. In the 1D distributions, the dashed lines mark the 90 per cent credible interval. The 2D plot shows contours of the 50 per cent and 90 per cent credible regions.}  
\label{fig:r4}
\end{center}
\end{figure}

The prior and posterior distribution of $\Delta \gamma$ for GW170104 and GW170823 is shown in Fig.~\ref{fig:r3}. We can see the most values of $|\Delta \gamma|$ are around the 0-axis, which may suggest WEP is valid on the intergalactic scale for GWs. In Fig.~\ref{fig:r2},  we can see that our results of SNR are similar to the LIGO-Virgo's for most GW events. While for GW151012, GW151226, GW170809, our SNRs are a little lower than LIGO ones. The influence of additional parameter $|\Delta \gamma|$ might be the reason. Most of the estimations of the BBH parameters coincide with LIGO-Virgo's results in the confidence interval. For example, Fig.~\ref{fig:r4} shows the posterior distribution of binary component masses and luminosity distance of GW170823,  which coincides to LIGO-Virgo's results.

\section{Discussion}
\label{sec:discussion}
Since the first direct detection of GWs\citep{abbott2016observation}, LIGO and Virgo collaboration has checked the consistency of several gravitational wave
 signals\citep{abbott2017gw170104, abbott2019tests} with the predictions of general relativity. By now, they give some constrains on modifications to the propagation of gravitational waves due to a modified dispersion relation, and they have not found any obvious inconsistency of the data with the predictions of GR \citep{abbott2019tests}. However, the WEP was still not tested with GW data analysis.

On the cosmology scale, by comparing the arrival time of GWs at different frequencies \citep{wu2016testing, wei2017multimessenger} or with Gamma-ray burst in GW170817, the deviation of WEP $\Delta \gamma$ has been constrained not larger than $\sim 10^{-9}$. In these tests, the observed time delay $\Delta t_{\mathrm{obs}}$ for two signals from the same source are directly treated as the one caused by WEP violation. The delay of the arrival times of two signals is completely attributed to the estimation of $\Delta \gamma$. These assumptions gave rough constraints on $\Delta \gamma$. This method could be improved by removing the intrinsic time delay if the physical mechanisms of the astrophysical events are known, such as GWs of the compact binary mergers. 

In this work, we study the intergalactic free-fall of GWs from all the BBH GW events in GWTC-1, which is LIGO-Virgo's first GW transient catalogue of compact binary mergers. Because the emission-time differences $\Delta t_{\rm{e}}$ of different frequency of GWs of one GW event are known exactly, any violation of WEP will contribute to the arrival time delay, and cause the dephasing of waveforms. This effect can be extracted in GW data. By considering the Shapiro time delay of the Milky Way's gravitational field, we construct a waveform template with $\Delta \gamma$. With this template, we analyse the GW data in GWTC-1, and constrain the $|\Delta \gamma| \lesssim 10^{-15}$ at 90 per cent confidence. This 
may imply that WEP is valid on the intergalactic scale for GWs.

There are two advantages in our model. First, the emission time delay of signals with different frequencies is known exactly. Second, the interstellar medium has no influence on the propagation of GWs. In addition, if we can include the gravitational potential of the host galaxies, the values of $\Delta \gamma$ should be constrained better. The relation of $\Delta \gamma$ and $\Delta f$ is also an interesting problem, which could be investigated in our future work. In the future, more and more GW events will be detected in the third observing run of LIGO-Virgo and future GW detections, we could expect that the constraint on WEP will be more stringent.

\section*{Acknowledgements}
This work was supported by NSFC(National Natural Science Foun- dation of China) no. 11773059, the Key Research Program of Frontier Sciences, CAS(Chinese Academy of Sciences), no. QYZDB-SSW- SYS016, MEXT(Ministry of Education, Culture, Sports, Science and Technology - Japan), JSPS(The Japan Society for the Promotion of Science) Leading-edge Research Infrastructure Program, JSPS Grant-in-Aid for Specially Promoted Research 26000005, JSPS Grant-in-Aid for Scientific Research on Innovative Areas 2905: JP17H06358, JP17H06361, and JP17H06364, JSPS Core-to-Core Program A. Advanced Research Networks, JSPS Grant-in-Aid for Scientific Research (S) 17H06133, the joint research program of the Institute for Cosmic Ray Research, University of Tokyo, the LIGO(The Laser Interferometer Gravitational-Wave Observatory) project, and the Virgo project. This work made use of the High Performance Computing Resource in the Core Facility for Advanced Research Computing at Shanghai Astronomical Observatory.
\section*{Data availability}

The data underlying this article will be shared on reasonable request to the corresponding author Wen-Biao Han.



\bibliographystyle{mnras}

\begin{thebibliography}{}
\makeatletter
\relax
\def\mn@urlcharsother{\let\do\@makeother \do\$\do\&\do\#\do\^\do\_\do\%\do\~}
\def\mn@doi{\begingroup\mn@urlcharsother \@ifnextchar [ {\mn@doi@}
  {\mn@doi@[]}}
\def\mn@doi@[#1]#2{\def\@tempa{#1}\ifx\@tempa\@empty \href
  {http://dx.doi.org/#2} {doi:#2}\else \href {http://dx.doi.org/#2} {#1}\fi
  \endgroup}
\def\mn@eprint#1#2{\mn@eprint@#1:#2::\@nil}
\def\mn@eprint@arXiv#1{\href {http://arxiv.org/abs/#1} {{\tt arXiv:#1}}}
\def\mn@eprint@dblp#1{\href {http://dblp.uni-trier.de/rec/bibtex/#1.xml}
  {dblp:#1}}
\def\mn@eprint@#1:#2:#3:#4\@nil{\def\@tempa {#1}\def\@tempb {#2}\def\@tempc
  {#3}\ifx \@tempc \@empty \let \@tempc \@tempb \let \@tempb \@tempa \fi \ifx
  \@tempb \@empty \def\@tempb {arXiv}\fi \@ifundefined
  {mn@eprint@\@tempb}{\@tempb:\@tempc}{\expandafter \expandafter \csname
  mn@eprint@\@tempb\endcsname \expandafter{\@tempc}}}

\bibitem[\protect\citeauthoryear{Abbott et~al.,}{Abbott
  et~al.}{2016}]{abbott2016observation}
Abbott B.~P.,  et~al., 2016, Phys. Rev. Lett., 116, 061102

\bibitem[\protect\citeauthoryear{Abbott et~al.,}{Abbott
  et~al.}{2017}]{abbott2017gw170104}
Abbott B.~P.,  et~al., 2017, Phys. Rev. Lett., 118, 221101

\bibitem[\protect\citeauthoryear{Abbott et~al.,}{Abbott
  et~al.}{2019a}]{abbott2019gwtc}
Abbott B.~P.,  et~al., 2019a, Phys. Rev. X, 9, 031040

\bibitem[\protect\citeauthoryear{Abbott et~al.,}{Abbott
  et~al.}{2019b}]{abbott2019tests}
Abbott B.~P.,  et~al., 2019b, Phys. Rev. D, 100, 104036

\bibitem[\protect\citeauthoryear{Ashton et~al.,}{Ashton
  et~al.}{2019}]{ashton2019bilby}
Ashton G.,  et~al., 2019, ApJS, 241, 27

\bibitem[\protect\citeauthoryear{Gao, Wu  \& M{\'e}sz{\'a}ros}{Gao
  et~al.}{2015}]{gao2015cosmic}
Gao H.,  Wu X.-F.,   M{\'e}sz{\'a}ros P.,  2015, ApJ, 810, 121

\bibitem[\protect\citeauthoryear{Hannam, Schmidt, Boh{\'e}, Haegel, Husa, Ohme,
  Pratten  \& P{\"u}rrer}{Hannam et~al.}{2014}]{hannam2014simple}
Hannam M.,  Schmidt P.,  Boh{\'e} A.,  Haegel L.,  Husa S.,  Ohme F.,  Pratten
  G.,   P{\"u}rrer M.,  2014, Phys. Rev. Lett., 113, 151101

\bibitem[\protect\citeauthoryear{Husa, Khan, Hannam, P{\"u}rrer, Ohme, Forteza
  \& Boh{\'e}}{Husa et~al.}{2016}]{husa2016frequency}
Husa S.,  Khan S.,  Hannam M.,  P{\"u}rrer M.,  Ohme F.,  Forteza X.~J.,
  Boh{\'e} A.,  2016, Phys. Rev. D, 93, 044006

\bibitem[\protect\citeauthoryear{Kahya \& Desai}{Kahya \&
  Desai}{2016}]{kahya2016djconstraints}
Kahya E.~O.,  Desai S.,  2016, Phys. Lett. B, 756, 265

\bibitem[\protect\citeauthoryear{Khan, Husa, Hannam, Ohme, P{\"u}rrer, Forteza
  \& Boh{\'e}}{Khan et~al.}{2016}]{khan2016frequency}
Khan S.,  Husa S.,  Hannam M.,  Ohme F.,  P{\"u}rrer M.,  Forteza X.~J.,
  Boh{\'e} A.,  2016, Phys. Rev. D, 93, 044007

\bibitem[\protect\citeauthoryear{Krauss \& Tremaine}{Krauss \&
  Tremaine}{1988}]{krauss1988test}
Krauss L.~M.,  Tremaine S.,  1988, Phys. Rev. Lett., 60, 176

\bibitem[\protect\citeauthoryear{Longo}{Longo}{1988}]{longo1988new}
Longo M.~J.,  1988, Phys. Rev. Lett., 60, 173

\bibitem[\protect\citeauthoryear{Mirshekari, Yunes  \& Will}{Mirshekari
  et~al.}{2012}]{mirshekari2012constraining}
Mirshekari S.,  Yunes N.,   Will C.~M.,  2012, Phys. Rev. D, 85, 024041

\bibitem[\protect\citeauthoryear{Shapiro}{Shapiro}{1964}]{shapiro1964fourth}
Shapiro I.~I.,  1964, Phys. Rev. Lett., 13, 789

\bibitem[\protect\citeauthoryear{Tingay \& Kaplan}{Tingay \&
  Kaplan}{2016}]{tingay2016limits}
Tingay S.~J.,  Kaplan D.~L.,  2016, ApJ, 820, L31

\bibitem[\protect\citeauthoryear{Touboul et~al.,}{Touboul
  et~al.}{2017}]{touboul2017microscope}
Touboul P.,  et~al., 2017, Phys. Rev. Lett., 119, 231101

\bibitem[\protect\citeauthoryear{Wei, Gao, Wu  \& M{\'e}sz{\'a}ros}{Wei
  et~al.}{2015}]{wei2015testing}
Wei J.-J.,  Gao H.,  Wu X.-F.,   M{\'e}sz{\'a}ros P.,  2015, Phys. Rev. Lett.,
  115, 261101

\bibitem[\protect\citeauthoryear{Wei, Wu, Gao  \& M{\'e}sz{\'a}ros}{Wei
  et~al.}{2016}]{wei2016limits}
Wei J.-J.,  Wu X.-F.,  Gao H.,   M{\'e}sz{\'a}ros P.,  2016, J. Cosmol.
  Astropart. Phys, 2016, 031

\bibitem[\protect\citeauthoryear{Wei et~al.,}{Wei
  et~al.}{2017}]{wei2017multimessenger}
Wei J.-J.,  et~al., 2017, J. Cosmol. Astropart. Phys

\bibitem[\protect\citeauthoryear{Will}{Will}{1998}]{will1998bounding}
Will C.~M.,  1998, Phys. Rev. D, 57, 2061

\bibitem[\protect\citeauthoryear{Will}{Will}{2014}]{will2014confrontation}
Will C.~M.,  2014, Living Rev. Relativ, 17, 4

\bibitem[\protect\citeauthoryear{Wu, Gao, Wei, M{\'e}sz{\'a}ros, Zhang, Dai,
  Zhang  \& Zhu}{Wu et~al.}{2016}]{wu2016testing}
Wu X.-F.,  Gao H.,  Wei J.-J.,  M{\'e}sz{\'a}ros P.,  Zhang B.,  Dai Z.-G.,
  Zhang S.-N.,   Zhu Z.-H.,  2016, Phys. Rev. D, 94, 024061

\bibitem[\protect\citeauthoryear{Xing, Gao, Wei, Li, Wang, Zhang, Wu  \&
  M{\'e}sz{\'a}ros}{Xing et~al.}{2019}]{xing2019limits}
Xing N.,  Gao H.,  Wei J.-J.,  Li Z.,  Wang W.,  Zhang B.,  Wu X.-F.,
  M{\'e}sz{\'a}ros P.,  2019, ApJ, 882, L13

\bibitem[\protect\citeauthoryear{Yao, Zhao, Han, Wang, Liu  \& Liu}{Yao
  et~al.}{2019}]{yao2019new}
Yao L.,  Zhao Z.,  Han Y.,  Wang J.,  Liu T.,   Liu M.,  2019, arXiv preprint
  arXiv:1909.04338

\makeatother
\end{thebibliography}




%
%


\bsp	
\label{lastpage}
\end{document}